\documentclass[%
 amsmath,amssymb,
 aps,
 prfluids,
]{revtex4-2}

\usepackage{graphicx}
\usepackage{tikz} 
\usepackage{dcolumn}
\usepackage{bm}
\usepackage{float}
\usepackage[normalem]{ulem}
\usepackage{xcolor}

\usepackage{pgfplots}

\usepackage{multirow}

\DeclareRobustCommand{\fcStar}[3]{
\begin{tikzpicture}[scale = (#1)/100]]
\draw
[draw=#2, line width=#3pt] 
(145,276) -- (190,380) --
(237,279) -- (349,288) --(278,199) -- (349,108) --
(237,121) -- (191,19) -- (147,118) -- (37,109) --
(98,198) -- (35, 287) -- (148,273);
\end{tikzpicture}
}

  \newcommand{\plq}[1]{\textcolor{teal}{#1}}

  \newcommand{\review}[1]{\textcolor{black}{#1}}
 \newcolumntype{M}[1]{>{\centering\arraybackslash}p{#1}}

\newcommand{\eg}{e.\,g.\ }

\begin{document}

\preprint{APS/123-QED}


 \title{From annular cavity to rotor-stator flow:  nonlinear dynamics of axisymmetric rolls}
 
\author{Artur Gesla\textsuperscript{\textdagger} }
 \altaffiliation[Also at ]{LISN-CNRS, Universit\'e Paris-Saclay, F-91400 Orsay, France \newline \hspace*{-7pt}\textsuperscript{\textdagger} now at École polytechnique fédérale de Lausanne, group HEAD }
\affiliation{Sorbonne Universit\'e, F-75005 Paris, France
}%


\author{Patrick Le Qu\'er\'e}
\author{Yohann Duguet}
\affiliation{LISN-CNRS, Universit\'e Paris-Saclay, F-91400 Orsay, France
}%
\author{Laurent Martin Witkowski}
\affiliation{%
 Universite Claude Bernard Lyon 1, CNRS, Ecole Centrale de Lyon, INSA Lyon, LMFA, UMR5509, 69622 Villeurbanne, France
}%


\date{\today}

\begin{abstract}

 Spatio-temporally complex flows are found at the onset of unsteadiness in (axisymmetric) rotor-stator turbulence in the shape of concentric rolls. The emergence of these rolls is rationalised using a homotopy approach, where the original flow configuration is continuously deformed into a simpler, better understood configuration. We deform here rotor-stator flow into an annular flow, thereby controlling curvature effects, and we investigate numerically the transition scenarios as functions of the Reynolds number. Increasing curvature starting from the planar limit reveals a clear path towards a subcritical scenario as a function of the Reynolds number. As the rotor-stator configuration is approached, supercritical branches shift to increasing Reynolds number while a subcritical branch of chaotic states takes over. Modal selection in the supercritical scenario involves the competition between two modal families. It rests on a specific radial localisation property of all eigenmodes, linked to the space-dependent convective radial velocity which intensifies as curvature is increased. A new nonlinear mechanism for the pairing of rolls is proposed based on multiple resonances. The critical point where the original rotor-stator flow loses its stability to axisymmetric perturbations is identified for the first time for the geometry under study.

\end{abstract}

\maketitle






     


\section{Introduction}

Owing to their relevance to many engineering configurations or geophysical situations rotating flows are one of the most investigated subjects of the fluid mechanics literature~\cite{Greenspan_1969, Pedlovski_1987, owen_1989a}. In standard geometries they are among the simplest academic flows 
used to illustrate
transition to unsteadiness. The main configuration of interest in this study is the {incompressible} flow between two disks, one rotating at a given angular speed, the other at rest. \review{If the disks extend to infinity, the governing equations admit similarity solutions, with the radial and azimuthal velocity components growing linearly with radius. The nonlinear system of differential equations has been shown to have non-unique solutions, at the origin of a notorious Stewartson-Batchelor controversy. The controversy was raised around the structure of the solution for large Reynolds number, made of a single boundary layer along the rotor for Stewartson~\cite{Stewartson_pcps_53}, as opposed to two separate boundary layers surrounding a uniformly rotating core for Batchelor~\cite{batchelor1951note}. Later, it was shown that the differential system in fact admits more solutions~\cite{Mellor1968flow,roberts1976computation,holodniok1977computation} characterized by their number of cells separated by lines of zero azimuthal velocity extending to infinity.} 
In the case of disks of finite radius bounded by an external shroud, which is the precise target configuration considered here, specific parameters are needed to describe the geometrical configuration, namely the ratio of the radial expanse to the distance between the disks, and also to describe the imposed boundary conditions. Early numerical studies \cite{dijkstra1983flow,adams1982incompressible,Szeri_jfm_83a,brady1987rotating,oliveira1991flow} have helped to understand the flow structure and the modifications induced by an external shroud. In particular the presence of an outer shroud has been shown to favor solutions \review{of Batchelor's type~\cite{batchelor1951note}}, although the core angular velocity is no longer independent of the distance to the rotation axis. This is particularly important for engineering applications since the local angular velocity of the core is directly linked to the radial pressure gradient and hence to the forces normal to the disks, a quantity of utmost interest for the design of bearings, for instance. Whether the shroud is attached to the rotor or to the stator also influences the flow structure. For a rotating shroud it was shown~\cite{Cousin_cras_99}, for one given value of $Re$, that the flow is made of three zones : a similarity zone close to the rotation axis of radial expanse approximately equal to the gap width, and {a second one in the form of} a recirculation zone close to the shroud, the expanse of which is also of the order of the gap {width}. {In between lies} a quasi-similarity zone, called {a} homothetic region, in which the velocity profiles considered at homothetic distances measured from a virtual origin of the axis coincide. 

From a fundamental viewpoint, we are interested in the bifurcations leading to an unsteady flow and, more generally, to turbulence, as the Reynolds number increases. \review{
Rotor-stator flows with moderate aspect ratio $\Gamma=R/H$ are then dominated by three dimensional structures, among which
spiral waves  \cite{launder2010laminar,martinand2023instabilities,xie2024onset}}.  
However, the first unsteady states appearing as $Re$ is increased are not spiral waves, but rather axisymmetric concentric rolls~\cite{schouveiler1999spiral,gauthier1999axisymmetric}, whose dynamics has 
become a subject of investigation 
\textit{per se}. 
These rolls have been observed first experimentally, then captured in simulations of finite cavities with 
$\Gamma=5$~\cite{serre2001annular,lopez2009crossflow} but only as short-time transients. For a cavity of aspect ratio
$\Gamma=10$, branches of subcritical axisymmetric nonlinear solutions were also identified in competition with a supercritical {axisymmetric} instability~\cite{daube2002numerical,gesla2024subcritical}, yet it was concluded that this branch of subcritical solutions does not extend  sufficiently low in $Re$ 
by comparison with those reported 
by experimentalists.  Moreover,
the rolls reported experimentally for these parameters 
\cite{schouveiler2001instabilities} 
are consistent with an input-output scenario, where parasitic vibrations of the set-up are selectively amplified by non-normal effects, leading to a response of the fluid system dominated by concentric rolls~\cite{gesla2024ontheorigin}. 
This input-output formalism
may give the impression that any response of the system is possible provided the right input.
Yet, independently of how they are forced the rolls seem to possess their own intrinsic dynamics, which still
remains to be understood. Experiments by Schouveiler~\cite{schouveiler1999spiral} show {indeed} a well-organised sequence of merging and pairing events as the propagation travels inwards along the stator boundary layer. This dynamics was first mentioned in numerical simulations~\cite{cousin1996thesis}. It was also reproduced numerically in~\cite{gesla2024ontheorigin} {in} the linear receptivity regime by imposing a multi-harmonic boundary forcing. Similar dynamics 
was
also reported for $\Gamma=5$ by Do {\it et al.}~\cite{do2010optimal} in the presence of a harmonic forcing only, this time \review{
attributed} to nonlinear interactions. \\

\begin{figure}
\includegraphics[width=0.5\linewidth]{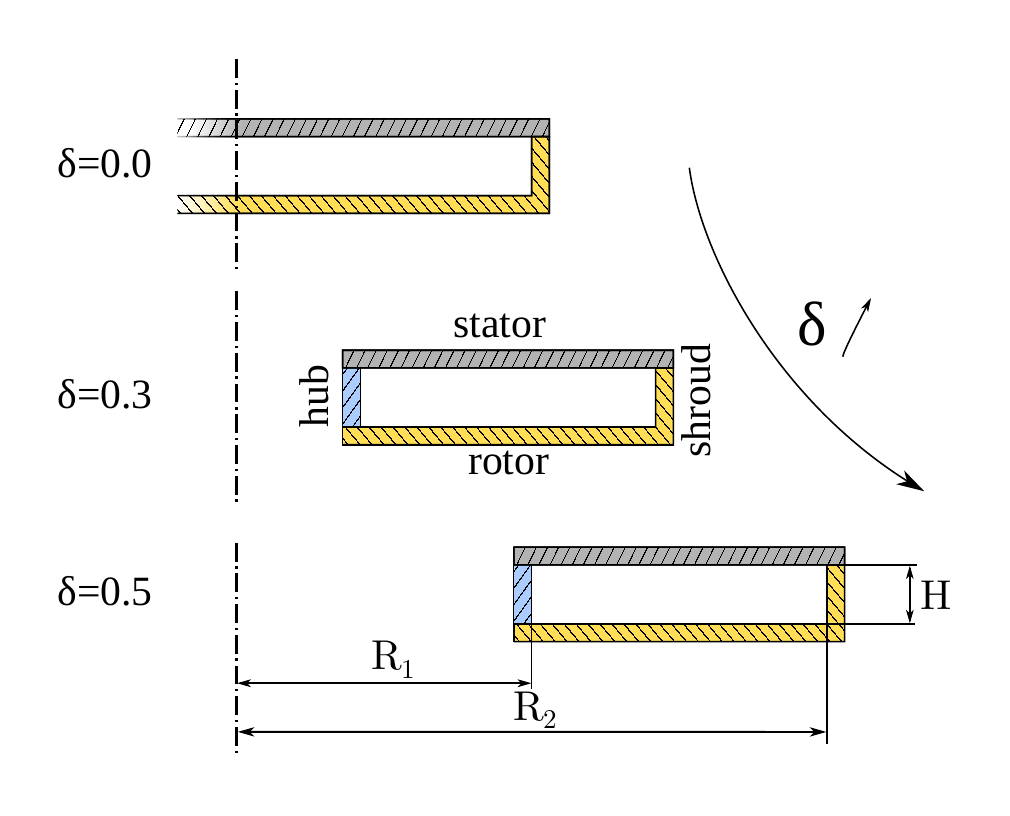}
\caption{Sketch of the system parametrised by $\delta=R_1/R_2$. 
$\delta \rightarrow 0$ : axisymmetric rotor-stator flow, $0<\delta<1$ : axisymmetric annular swirling flow, $\delta \rightarrow 1$ : parallel Ekman boundary layer flow analog to a two-dimensional differentially heated cavity flow (DHC).}
\label{fig.1}
\end{figure}

Rotor-stator  cavities with $\Gamma=5$ have been extensively studied in the past as a compromise between radial extension and numerical requirements~\cite{serre2001annular, serre_pof_2004, Tuliszka_jtam_2005,Severac_pof_2007,poncet2009revisiting, lopez2009crossflow, do2010optimal,  makino2015laminar, bridel-bertomeu2017large, 
queguineur_pof_2019}.
Several variations have been considered depending whether the cavity extends to the axis, (in which case it is referred to as a cylindrical cavity), or whether it features an inner hub or shaft (in which case the cavity is referred to as annular). 
Cylindrical cavities are characterized by a single geometrical parameter, their aspect ratio $\frac{R_2}{H}$, whereas annular cavities require an additional geometrical parameter to characterize the 
curvature, 
\review{which we chose as $\delta = \frac{R_1}{R_2}$ (see  Fig.~\ref{fig.1})}.
\
Note also that additional variability comes from the nature of the boundary conditions for the velocity on the inner hub and outer shroud. The shroud has generally been assumed 
\review{
}
attached either to the rotor (\eg \cite{Lopez_pof_1996,Daube_cf_2002,Lopez_pof_2009}) or to the stator (\eg \cite{Itoh_ETSF_1992,Gauthier_jfm_1999,Schouveiler_jfm_2001,Severac_pof_2007,Severac_pof_2007,Poncet_pof_2009,Makino_ftc_2015}). 
\review{This} \review{has raised the question of the numerical treatment of the corner singularity.
Various strategies have been considered either to mimic the experiments~\cite{dijkstra1983flow} 
or to alleviate the feared detrimental consequences of this corner singularity 
on the accuracy and dynamics of numerical solutions, in particular when using Chebyshev spectral methods~\cite{Lopez_jcp_1998,serre2001annular}. 
Others have used singular-free albeit non-physical linear velocity profiles 
for the azimuthal velocity~\cite{cousin1996thesis,cousin1998nature,serre_jfm_2001}. All these specific features, 
beside introducing additional parameters, have non-negligible effects on the flow dynamics and its stability  characteristics,
effects which have not been investigated systematically.}
\\

The aim of the present study is to cast light on the intrinsic dynamics of the concentric rolls. The present approach is based on the idea of homotopy where the system of interest is considered part of a larger family of flow configurations. 
In homotopy studies of fluid flows, a given flow configuration is deformed continuously into another potentially simpler one, by either varying a single geometric parameter $\delta$ or by adding an external body force \cite{nagata1990three,clever1992three,faisst2000transition}. Homotopy methods can be used either for the detailed continuation of exact nonlinear solutions between two different parameter values or, more loosely, as a way to track turbulent regimes as a function of a given additional parameter \cite{ishida2017turbulent}.
In the present study, 
restricted to an (axisymmetric) rotor-stator cavity of radial aspect ratio $\Gamma=5$, the cavity is deformed continuously by considering a central hub of growing radius {while keeping $\Gamma$ constant},
 as sketched in Fig.~\ref{fig.1}. 
 The part of the domain where the fluid flow occurs can thus be pushed to infinity where curvature
effects are expected to vanish.
 Stress-free boundary conditions are imposed at the hub in order to make this homotopy concept consistent for {the rotor-stator limit}
 $\delta \rightarrow 0$. \\



\review{As it turns out, 
inspired by
the well known analogy between rotation and stratification mentioned by Veronis~\cite{veronis1970analogy}, the flow in a rotor stator cavity
bears a close resemblance with the flow in a differentially heated cavity (DHC), the azimuthal velocity playing the role of the temperature for a Prandtl number is unity. An adiabatic condition for temperature translates into a stress-free boundary condition for the azimuthal velocity. The exact conditions for this analogy \review{depend on several independent parameters, such as differential rotation and curvature, and} are not discussed here in detail.  A similar analogy between Taylor-Couette and Rayleigh-Bénard configuration is documented by \cite{Chandrasekhar_61,eckhardt2020exact}.} 
 %
\review{The benefit behind the present analogy is that the steady flow in the DHC with unity Prandtl number always lose stability and become unsteady at finite parameter values through well understood, supercritical linear instabilities~\cite{le1998onset,henkes1990,Xin_pof_01,gadoin_ijnmf_2001,xin2002extended,xin2006natural,xin2012stability,oteski2015quasiperiodic,net2017periodic,khoubani2024vertical}. Analogous destabilisation mechanisms are expected to take place in the high-$\delta$ rotor-stator configuration, leading to non-trivial flow solutions that can be easily identified. Such solutions can then be followed by straightforward continuation down to lower values of $\delta$ where they would be otherwise difficult to identify.}
\review{ Even if, experimentally, three-dimensional structures 
would dominate the flow for most parameters of this study, we chose to analyse only the axisymmetric flow. Restricting this study to an axisymmetric rotor-stator configuration is
only possible in numerical simulation.  We use the axisymmetry assumption
in a constructive manner by focusing exclusively on the axisymmetric mechanisms responsible for the observation of the concentric rolls, independently of their interaction with non-axisymmetric modes.}
\\

{The present article is structured as follows. In Section II, we review the mathematical formulation of the problem and introduce all the parameters relevant to numerical simulation and to the homotopy concept. Section III contains the results of linear stability analysis, including a spatial description of the base flow, of the leading eigenmodes and an estimation of non-normal amplification by the base flow. In Section IV, instances of nonlinear dynamics observed as $\delta$ is lowered are reported, including an original dynamical interpretation of the pairing of rolls. Finally, in Section V, the path towards the rotor-stator configuration $\delta=0$
is described based on the
numerical exploration. Conclusions and outlooks are given in Section VI.}\\

\section{Equations and algorithms}


\subsection{Formulation}

We consider the incompressible flow of a Newtonian fluid with kinematic viscosity $\nu$ inside an annular cavity formed by a hub, a shroud and two {parallel} disks. The geometry features an axial gap $H$, an inner radius $R_1$ and an outer radius $R_2${, as} sketched in Figure 1. The fluid motion is induced by the rotation at constant angular velocity $\Omega$ of one of the two disks (the bottom one in Figure 1), the other being kept fixed. We
use $H$ as the reference length scale and $\Omega R_2$ as the reference velocity scale, and define the Reynolds number accordingly as $Re=\Omega \, H \, R_2/\nu$. The aspect ratio of the cavity is defined as $\Gamma=(R_2-R_1)/H$
and the radius ratio (our main homotopy parameter) as $\delta=R_1/R_2$.  We use the non-dimensional radial coordinate $r$ going from the hub at $r_1=R_1/H = \frac{\Gamma \; \delta}{1 - \delta}$ to the shroud at $r_2=R_2/H  = \frac{\Gamma }{1 - \delta}$, while the non-dimensional axial variable $z$ goes from the rotor at $z=0$ to the stator at $z=1$. Time is expressed in units of $H/\Omega R_2$.
Throughout the whole paper we will stick to the case $\Gamma=5$, while also referring to the results obtained for $\Gamma=10$ and $\delta=0$ in~\cite{daube2002numerical}, revisited recently in~\cite{gesla2024subcritical,Gesla_PhD_2024}.
Under the assumption that the flow 
is axisymmetric and incompressible, the non-dimensional governing equations are


\begin{subequations} \label{ns}
\begin{align}
\frac{\partial \mathbf{u}}{\partial t}+
\nabla( \mathbf{u} \otimes \mathbf{u})
&=-  \mathbf{\nabla} p + \frac{1}{Re} \mathbf{\nabla}^2  \mathbf{u} \\
 \mathbf{\nabla} \cdot  \mathbf{u}&=0
\end{align}
\end{subequations}
where $\otimes$ denotes the tensorial product and
$p$ is the non-dimensional pressure field. We assume that the disk at $z = 0$ and the shroud rotate at the same speed. The no-slip boundary conditions on the disks and on the shroud are
\begin{eqnarray}
	{\bm u} & = & {\bm 0} \mbox{ at } z=1 \nonumber  \\
	{\bm u} & = & H \,r/R_2 \,{\bm e_{\theta}} \mbox{ at } z=0 \nonumber  \\
	{\bm u} & = & {\bm e_{\theta}} \mbox { at } r_2=R_2/H, \nonumber  
\end{eqnarray}
with ${\bm e_{\theta}}$ the unit azimuthal basis vector. 
At $r_1=R_1/H$ we use the stress-free condition
\begin{equation} 
\label{BCR2}
( u_r \,, \frac{\partial u_{\theta}}{\partial  r}-\frac{u_{\theta}}{r} \,, \frac{\partial u_{z}}{\partial r}) = (0, 0, 0) \mbox { at } r_1=R_1/H.
\end{equation}
In the specific case $\delta \rightarrow 0$, the boundary condition at $r_1=R_1/H$ becomes a condition at the axis $r=0$, which reads \begin{eqnarray}
(u_r \,, u_{\theta} \,, \frac{\partial u_{z}}{\partial r}) = (0, 0, 0).
\end{eqnarray}

\
For increasing $\delta$ the curvature effects in the cavity decrease and at the limiting \emph{planar} asymptote $\delta \rightarrow 1$
the system {ruled by Eq.
\ref{ns}}
becomes
analogous 
to the two-dimensional flow of a virtual heat-conducting fluid with Prandtl number unity, occurring inside a rectangular cavity of  
{
vertical aspect ratio  $\Gamma$, \review{as mentioned in the introduction}.} 

\subsection{Methodology}

{Computational} results were obtained with two main tools : time integration of the governing equations or steady state solving  and computation of corresponding eigenmodes.
The unsteady equations written for the velocity perturbations as primary variables are integrated in time using the classical prediction-projection fractional step algorithm \cite{Guermond_cmame_2006}. The spatial discretisation is of finite volume {type} on staggered grids, {with $N_r$ and $N_z$ respectively the number of internal pressure cells in $r$ and $z$.} The prediction step computes a provisional velocity field  using the second order three time-level scheme, which combines an implicit treatment of the diffusion terms using Backward Difference Formula 2 with an explicit Adams-Bashforth 2 discretisation of the convection terms~\cite{VanelPeyret_1986}. 
This velocity field is then projected onto the space of divergence-free vector fields to yield the final velocity field. The projection step relies on solving an elliptic equation for the correction pressure. This is performed in practice by direct methods ensuring that the divergence of the velocity field at the end of the each time step is at machine accuracy, in order to avoid possible artefacts due to residual errors of iterative methods. 
Steady state solutions were obtained using Newton's iterations via the use of MUMPS~\cite{Amestoy_siamjmaa_2001} to solve linear systems, 
with the Jacobian matrix {stored} explicitly in compressed format.

The stability of the steady solutions was determined by computing their {eigenspectrum using} ARPACK~\cite{Arpack_98} in shift-and-invert mode, requiring a few tens of eigenmodes for each value of the shift.
Note also that the unsteady equations are integrated in incremental form  \cite{Gesla_PhD_2024}. This choice allows, when starting from a steady solution  converged to machine accuracy using the Newton scheme, to avoid the large transient growth issues due to non-normality, that would occur if using a non-incremental form. This is due to the right hand side of \review{
time discretised momentum equations} 
{\review{in the prediction step}} being exactly, at the first timestep, the steady state residual of the last Newton iteration. \review{Details of this implementation are available in \cite{Gesla_PhD_2024}.}\\



\subsection{Numerical requirements}


As forthcoming results will show, simulating numerically this configuration near criticality becomes increasingly demanding as curvature effects grow. This difficulty culminates with the cylindrical rotor-stator configuration. 
A grid sensitivity test reported in Table~\ref{tab:my_label} of the Appendix A shows that the leading eigenvalue responsible of the loss of stability of the base flow is very sensitive to the numerical resolution while the others are not (see fig.\ref{fig:my_label}). This confirms the observation made in ref.\cite{gesla2024subcritical, Gesla_PhD_2024} for a rotor-stator cavity with $\Gamma = 10$.
This test demonstrates that it is computationally demanding to claim spatial convergence with reasonable computing resources, whatever the spatial \plq{
\review{discretisation}} used.
Visual inspection reveals that the main difficulty deals with capturing the small vortical structures, for instance present
in the regular corner at 
\review{$ (r,z )= (r_2,0) $}.
As a compromise, all our  investigations are carried out for a nominal resolution $N_r \times N_z = 512\times 256$ with a uniform grid in $r$ and a cosine grid in $z$.\\



\section{Base flow, eigenmodes and linearised dynamics }


\subsection{Base flow}

\begin{figure}
    \centering
    \includegraphics[width=\linewidth]{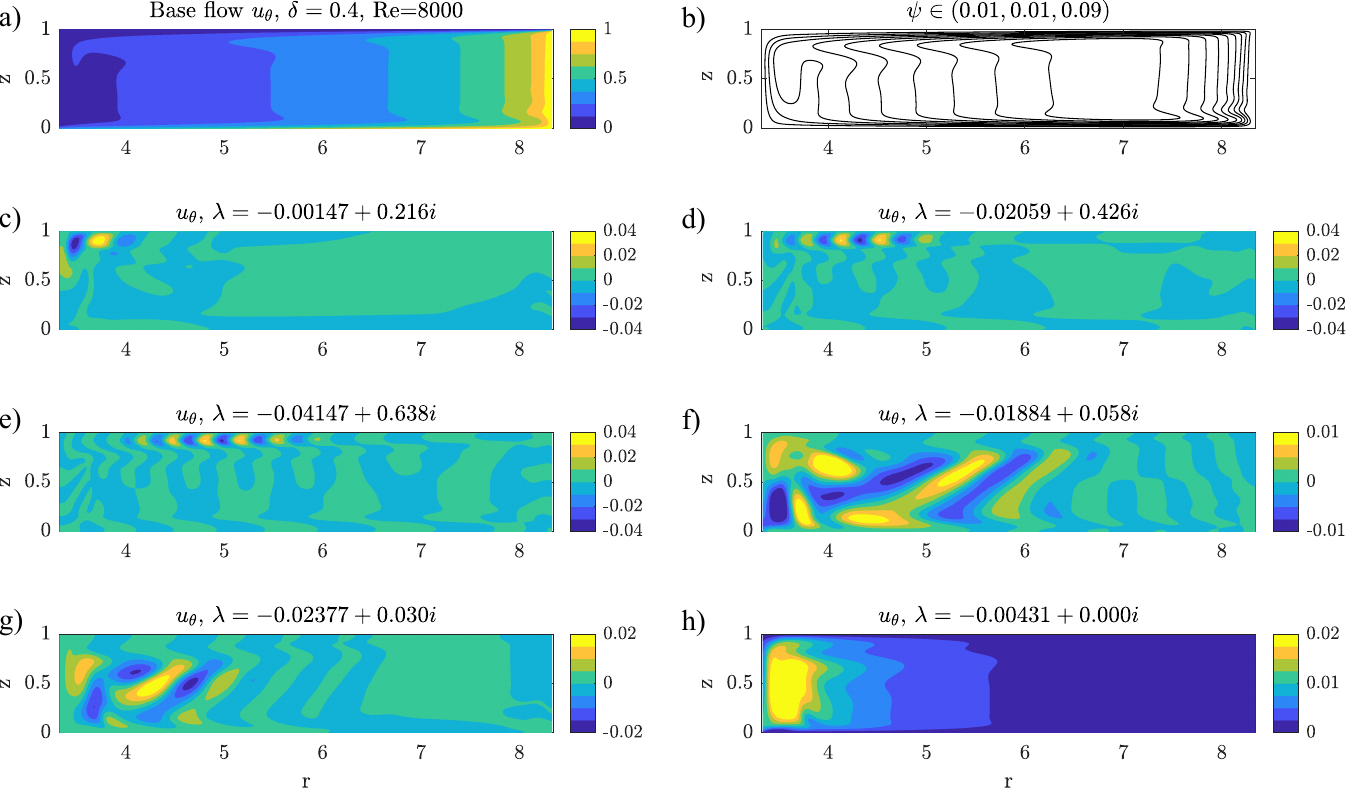}
    \caption{
    Base flow and related eigenvectors for $Re=8000$ and $\delta=0.4$. a) Azimuthal velocity $u_{\theta}(r,z)$ of the base flow  b) corresponding meridional streamlines c) Azimuthal velocity $u_{\theta}(r,z)$ of the least stable eigenvector d)-h) other eigenvectors. In f) and g) inertial waves are visible.
    }
    \label{fig.2}
\end{figure}

For each parameter $Re$ and $\delta$ the system of equations admits a steady axisymmetric solution denoted as the \emph{base flow}. It is determined up to machine accuracy using the same standard Newton--Raphson solver as in Ref.~\cite{Faugaret_jfm_2020}. The associated velocity field ${\bm U_b}(r,z)$ is axisymmetric but features three components. Its structure is displayed in figure~\ref{fig.2}(a,b). It combines a shear azimuthal flow with a recirculating meridional flow. For large enough $Re$ this meridional flow consists of a parallel Ekman layer next to the rotor and a spatially developing B\"odewadt layer on the stator, with the flow directed respectively outwards and inwards.\\

\subsection{Critical Reynolds number \review{and eigenmodes}}

\begin{figure}
\centering
\includegraphics[width=0.5\linewidth]{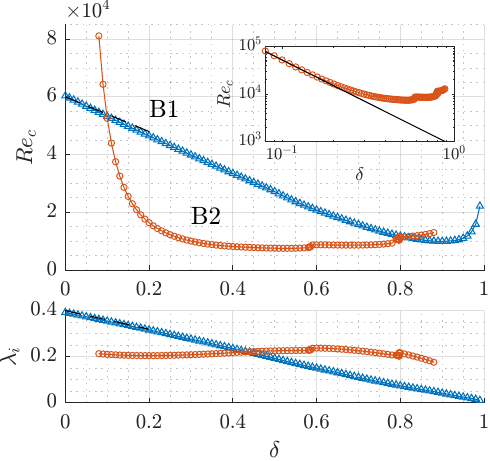}

\caption{Critical Reynolds number $Re_c$ (top) and corresponding marginal eigenfrequency $\lambda_i$ (bottom) depending on the curvature parameter $\delta$. Branch B2 asymptotes to $Re_c=O(\delta^{-a})$ for $\delta \rightarrow 0$, with $a \approx 2$,
which leads to a diverging threshold for the corresponding eigenmodes. On the other hand, continuation of the branch B1 leads to a finite value of $Re_c \approx 6 \times 10^4$ for $\delta=0$, the corresponding eigenmode being shown in Fig.~\ref{fig:B1}. Dashed lines plotted for $\delta<0.2$ and overlapping with branch B1 correspond to constant values of \review{$Re\,r_2 = 3 \times10^5$} and \review{$\lambda_i\,r_2=2$}. 
}
\label{fig_rec}
\end{figure}

The linear stability of this base flow can be monitored using \plq{\review{ARPACK in }}the shift-and-invert {mode} 
mentioned earlier. The eigenvalue problem involves
an \emph{ansatz} of the form ${\bm u} \sim e^{\lambda t}$, with $\lambda=\lambda_r+i\lambda_i$, so that a positive real part $\lambda_r>0$ indicates instability. A critical Reynolds number $Re_c$ can be defined where the real part of an eigenvalue 
vanishes. The critical values of $Re$ associated with the two dominant eigenvalues are shown in 
fig.~\ref{fig_rec}\review{(top)} as a function of $\delta$ ranging from 0 to 1, with key values listed in Table~\ref{tab:1}. The corresponding imaginary part is reported in fig.~\ref{fig_rec}\review{ (bottom),} which allows one to distinguish branches. The family of eigenmodes labelled $B2$ is
most unstable for all $\delta's$ between $\delta_1\approx0.1$ and $\delta_2\approx0.8$.
Several points of codimension two, not studied in detail here, are visible as kinks along the branch. They suggest that branch $B2$ is not governed by a single individual branch of eigenvalues but rather consists of several of them. Near $\delta \rightarrow 0$, 
the data is consistent with a divergence of $Re_c$ as a power law of the form $Re_c=O(\delta^{-a})$ with $a \approx 2$.
Between $\delta=0$ and $\delta<\delta_1$ another family of modes labelled $B1$ takes over as the most unstable. The dependence of $Re_c$ is affine in $\delta$, 
 which leads to a finite value of $Re_c$ at $\delta=0$. In the interval $\delta\in(0,\delta_1)$ the product $Re_c \,r_2$ is almost constant and equal to $3\times 10^5$. It is noted that Ref.~\cite{gesla2024subcritical} reports the value of $Re_c \,r_2=2.93\times10^5$ for {another} aspect ratio {of} 10, hence this result may hold for intermediate values of  $\Gamma$ too.
  This finite threshold {is reported to our knowledge for the first time. In particular, it was not reported in  previous investigations of rotor-stator flow with aspect ratio 5 in Ref.~\cite{lopez2009crossflow}.} 
 The branch B1 taking over between $\delta=\delta_2$ and $\delta=1$ supports frequencies vanishing linearly as 
 $\delta \rightarrow 1$ ({we recall that time is in units of $H/R_2\Omega$}). As $\delta \rightarrow 1$, the corresponding $Re_c$ diverges, at least using the current definition of $Re$.
\\



\begin{table}
    \centering
    \begin{tabular}{c|ccccccccccc}
$\delta$   & 0.00   & \textbf{0.10}   & 0.20   & 0.30   & 0.40   & 0.50   & 0.60   & 0.70   & \textbf{0.80}   & 0.90   & 0.99   \\ \hline 
${R_1}/{H}$ & 0. & 0.55 & 1.25 & 2.14 & 3.33 & 5. & 7.5 & 11.67 & 20. & 45. & 495. \\
${R_2}/{H}$ & 5. & 5.55 & 6.25 & 7.14 & 8.33 & 10. & 12.5 & 16.67 & 25. & 50. & 500. \\
$Re_c$   & {59564}   & \textbf{52573}   & 16408   & 10054   & 8165   & 7572   & 8764   & 8646  & \textbf{11462}   & 9960   & 22155   \\  
$\lambda_i$   & 0.392   & \textbf{0.211}   & 0.205   & 0.209   & 0.218   & 0.226   & 0.238   & 0.226   & \textbf{0.216}   & 0.042   & 0.011   \\  
 
    \end{tabular}
    \caption{Values of non-dimensional inner and outer radius, critical Reynolds number $Re_c$ and corresponding marginal eigenfrequency $\lambda_i$ depending on the curvature parameter $\delta$. Approximate codimension-2 points are marked in bold. A sharp decrease of $\lambda_i$ at $\delta=0.1$ and $0.8$ corresponds to the switch from B1$\rightarrow$B2 and B2$\rightarrow$B1 respectively.  }
    \label{tab:1}
\end{table}

\begin{figure}
    \centering
    
    
    
    
    
    \includegraphics[width=0.5\linewidth]{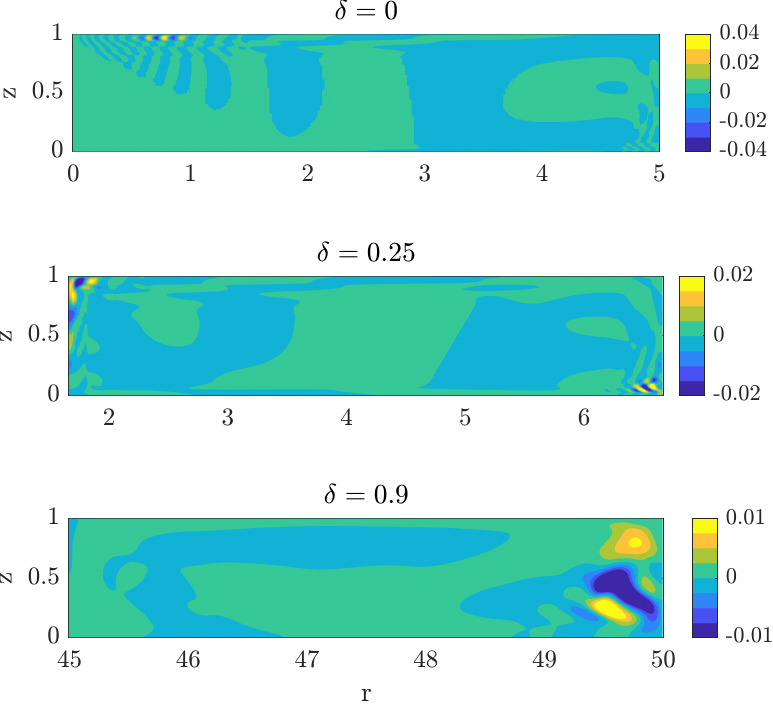}
    \caption{Azimuthal velocity of the neutrally stable eigenvector on the branch B1. Structure of the eigenvector evolves gradually from the rolls in the B\"odewadt layer for $\delta=0$, through the structures of similar amplitude in two opposite corners of the cavity for $\delta=0.25$, to the corner structure near the shroud $\delta=0.9$. For the corresponding values of $Re_c$ and $\lambda_i$ see figure \ref{fig_rec}. 
    }
    \label{fig:B1}
\end{figure}

The azimuthal velocity of a few representative selected eigenmodes is shown in fig.~\ref{fig.2}. Azimuthal velocity fields typical of the eigenvectors from  branch $B1$ are shown in fig.~\ref{fig:B1}. 
The eigenmodes can grossly be classified into distinct categories depending on the spatial region where the velocity displays the largest amplitude : 
\begin{itemize}
    \item \emph{B\"odewadt modes} These modes are made of counter-rotating vortices located strictly inside the B\"odewadt layer along the stator. Such modes have a non-zero phase velocity oriented inwards towards the rotation axis, hence they correspond to boundary layer modes propagating in the direction of the flow. The B\"odewadt boundary layer grows spatially
    {and}
    the eigenfrequencies depend on the spatial localisation of the mode, as will be detailed in the next section. 
    Such modes are shown in fig. \ref{fig:B1}\review{(top)} and also in fig. \ref{fig.2}(c,d,e), with three frequencies forming a triad of the form {$f_1= \pm f_2 \pm f_3$}
    \item \emph{Shift modes 
    } These steady modes are located in the bulk near the inner boundary at $r=R_1/H$, there where the velocity field was reported as self-similar~\cite{Cousin_cras_99}. Their importance has been noted for generic Hopf bifurcations~\cite{Noack2003hierarchy}, as such modes are in connection to the second order mean flow modification induced by self-interaction of the primary Hopf mode.
    Such a shift mode is shown in fig. \ref{fig.2}(h) 
    \item \emph{Shroud modes} 
    They correspond to streaks attached to the shroud. An example of a shroud mode is shown in fig.~\ref{fig:B1}\review{(bottom)}. This corner structure is analogous to the first instability mode \review{found} in low aspect ratio \review{DHC} cavities with adiabatic top and bottom walls \cite{le1998onset,xin2002extended,xin2006natural,xin2012stability,oteski2015quasiperiodic,khoubani2024vertical}
    \item \emph{Inertial bulk modes} 
    These modes are located exclusively in the bulk but unlike shift modes they also occur away from the rotation axis. These modes are propagating modes. They feature a patterned spatial structure with several clearly oblique wave vectors varying with radial location. Since the local rotation varies with the radius, these eigenmodes are labelled \emph{inertial} by an incomplete analogy with inertial modes in a rotating fluid~\cite{Greenspan_1969}, however with locally varying angular speed~\cite{fabre2006kelvin}. When the rotation rates is spatially constant, inertial waves are known to propagate along directions making a constant angle with the rotation axis that depends only on their frequency. Large-scale gradients of rotation rate are expected to distort the wave pattern in such a way that for a given frequency the angle depends now on position. This is precisely observed in fig.~\ref{fig.2}(f), most clearly for $4\le r\le7$ where the angle increases as the core rotation rate decreases with decreasing radius. 
\end{itemize}

The topology and the classification  of the modes of the branch $B1$ morphs, as $\delta$ moves from 
values close to 1 down to 0, from shroud-localised modes to B\"odewadt roll-like modes combined with inertial oscillations emitted from the roll part. \\

Unlike what fig. \ref{fig_rec} could suggest, the competition does not involve solely $B1$ and $B2$.
B1 was traced to show that it is a continuation of the same eigenmode which is most unstable at both end regions of variation of $\delta$, although, as fig.~\ref{fig:B1} shows, there seems to be little in common between the eigenmode structure at both ends. When B2 is first unstable, that is for $ 0.1 \lesssim \delta \lesssim 0.8 $, there are often many modes that become unstable before $Re$ is increased up to the value  corresponding to the B1 curve. For instance, for $\delta = 0.4$, the evolution of the spectrum in fig. \ref{fig:rzif} shows that for $ Re = 2 \times 10^4$, 
there are as many as 15 other eigenmodes which are linearly unstable. This situation is common for all values of $\delta > 0.3$. For  $ \delta < 0.2$, however, B1 and B2 modes are the first ones to become unstable. \\

As mentioned in the introduction, additional variability comes from the nature of the boundary conditions for  the  velocity  on  the  inner  hub  and  outer  shroud. The values of $Re_c$ are documented, for the case of $\delta=0.5$, for various types of boundary conditions. The comparison is shown in Table \ref{table:bcs} in Appendix B.  \\


\subsection{Radial localisation}

\begin{figure}
    \centering
        \includegraphics[width=0.49\linewidth]{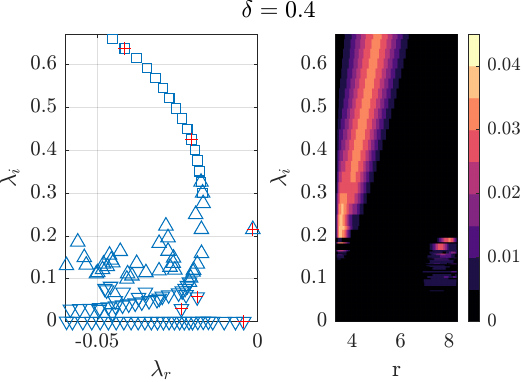}
        \includegraphics[width=0.49\linewidth]{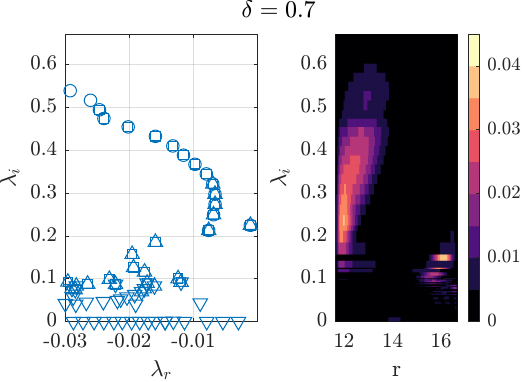}
    \caption{Spectrum of linearised Navier--Stokes operator and radial support of all (normalised) eigenmodes parametrised by their eigenfrequency $\lambda_i$ (amplitude of $u_{\theta}(z=z^*)$ for $z^*=0.94$
    ) at Re=8000 for $\delta=0.4$ (a) and at Re=8500 for $\delta=0.7$. The center of mass of the spatial support varies linearly with $\lambda_i$ (for $\lambda_i>0.2$). Eigenvectors from figure \ref{fig.2} correspond to red crosses in (a). Spectrum is obtained using series of shifts in Arnoldi method and plotted using different symbols. 
    }
    \label{fig:spectra}
\end{figure}

When computing the eigenspectrum of the system linearised around the steady base flow solution, an original property emerged for all $\delta < 1$. It is best seen in Fig.~\ref{fig:spectra} where
 eigenvectors, associated with the eigenvalues of largest real part, are plotted 
 in the plane $(r,\lambda_i)$, where $\lambda_i$ is the imaginary part of a complex eigenvalue $\lambda$. More precisely, the 
 kinetic energy of each eigenvector, which is independent of its phase, is considered at axial position $z=z^*$ with $z^*=0.94$ 
 (next to the stator) and plotted radially.
 In this $(r,\lambda_i)$ representation, each eigenspectrum appears as a coherent  rectilinear beam-like shape stretching from the hub towards the shroud as $\lambda_i$ increases.\\
 


\begin{figure}
\centering

    \includegraphics[width=0.45\textwidth]{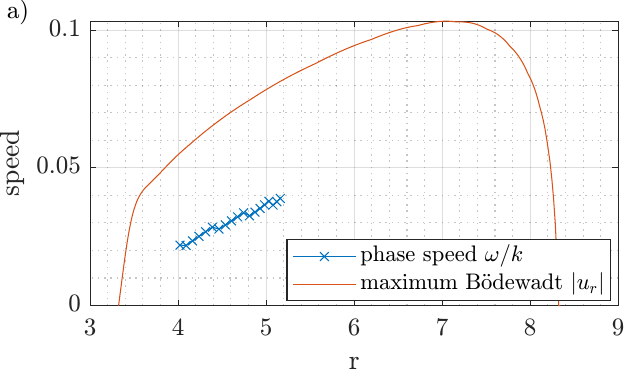}
 \hspace*{0.07\textwidth}
    \includegraphics[width=0.45\textwidth]{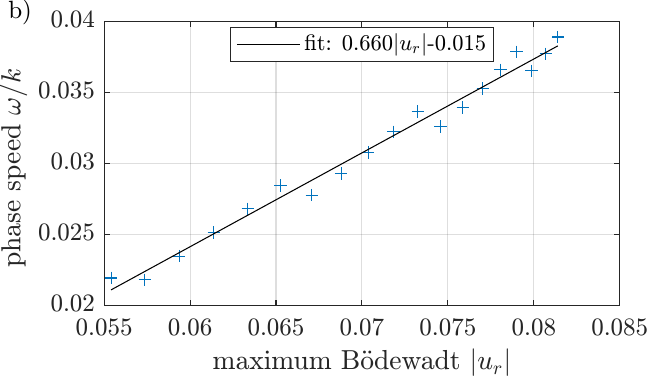}
    \caption{Comparison of the maximal radial velocity associated to the base flow and the 
    phase speed of the waves corresponding to the eigenvectors of the base flow (cf. fig.~\ref{fig:spectra}) for Re=8000 and $\delta=0.4$. In (a) both phase speed and a local radial base flow velocity are plotted as the function of $r$. In (b) the linear fit shows that in the current configuration the wave speed is a linear function of the local maximal radial base flow velocity equal $40\%\ -\ 50\%$ of the local maximal radial base flow velocity. }
    \label{fig:phase-speed}
\end{figure}

The direct interpretation is that all eigenvectors with spatial support localised inside the B\"odewadt layer see their active region localised radially at a position depending only on the value of the angular frequency $\lambda_i$. This makes radial localisation the signature of a given eigenmode (or rather of its natural oscillation frequency). {For a subset of eigenvectors from fig.~\ref{fig:spectra} corresponding to $\lambda_i>0.3$ the phase speed of the waves constituting the eigenvector was computed as the ratio between the oscillation angular frequency $\omega$ and the local radial wave number $k$ obtained from a spatial Fourier transform in $r$ at $z=z^*$. 
\review{The phase number associated with the localised wavepacket can be efficiently computed using a Fourier transform 
of the whole radial sample at a given $z$. The wavenumber corresponding to the maximal amplitude of the Fourier spectrum was associated with the wavenumber characterizing the wavepacket. }
This phase speed is compared against the local maximal radial speed in the B\"odewadt layer in figure \ref{fig:phase-speed}, showing an {unambiguous} linear dependence between {them}.} This correlation between wave speed and radial velocity is more marked when $\delta$ is far from 1.
This is consistent with the fact that for $\delta \rightarrow 1$ 
curvature vanishes,
and hence no large scale radial velocity gradient.

\subsection{Non-normal amplification}

Linear stability analysis of the base flow, via the listing of unstable eigenvalues, represents only partial knowledge of the 
stability of the problem. As is well known from investigations of shear flows, the finite-time stability of the system with respect to infinitesimal perturbations to the base flow is also of interest. In particular, if the operator resulting from the linearised equations is non-normal (i.e. it does not commute with its adjoint), finite-time amplification of the perturbation energy $E(t)=\int |{\bm u'}|^2d^3x$ is possible, where ${\bm u'}={\bm u}-{\bm U_b}$. \\

\begin{figure}
    \centering
    \includegraphics[width=0.5\linewidth]{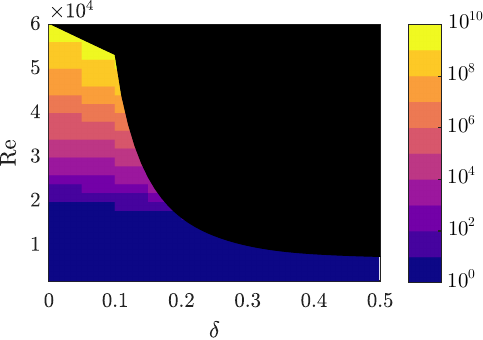}
    \caption{Energy gain $G$ in logarithmic scale as a function of $\delta$ and $Re$ (see eq.\eqref{gain}), where G is the maximal energy amplification by the linearised dynamics for a given random perturbation. For low $\delta$, large gain values are a consequence of the non-normality of the underlying linear operator. The part of the ($\delta,Re$) space corresponding to a linearly unstable base flow (cf. fig.~\ref{fig_rec}) is filled with black. 
    }
    \label{fig:logG}
\end{figure}

In principle, optimal growth can be investigated using the toolbox presented in ref.~\cite{schmid2007nonmodal}. We preferred the more economical
non-optimal approach of i) considering one generic enough initial condition and ii) monitoring the energy amplification associated with that initial condition (computed using standard time-stepping of the linearised equations). This is achieved using an initial condition $u'_{\theta}$ consisting of a {random field of small amplitude}. The energy gain 
\begin{equation} \label{gain}
    G(t)=\max_{t\ge 0} \frac{E(t)}{E(0)}
\end{equation}
is plotted in fig. \ref{fig:logG} \emph{versus} $Re$ and $\delta$, focusing only on the values of $\delta \le 0.5$.
Fig.~\ref{fig:logG} shows 
that $G$ increases fast with $Re$. For each individual value of $\delta$, the data is compatible with a fit of the form $G=e^{aRe}$  \review{in connection with the results from \cite{gesla2024ontheorigin}.} \review{ This scaling
was already reported by \cite{xu2021non,moron2022effect} in pulsatile pipe flow, where it
was justified by focusing on the transient part of the time interval during which the laminar profile is (locally) linearly unstable. Here, similarly, optimal perturbations travel downstream through regions characterised by either local stability or local instability, hence the time spent in the locally unstable region might justify the exponential scaling observed for the energy gain. This question deserves deeper analysis.} {Between $\delta=0.5$ and $\delta=0$}, the value of $G$ at $Re=Re_c$ increases by almost 10 decades. 
If we follow the neutral curve $Re_c(\delta)$  towards $\delta=0$, the global tendency is a sharp rise of $G$ especially for $\delta \le 0.15$.
Strong non-normal growth implies rapid \review{
transient} divergence between state space trajectories. In particular, even if a stable supercritical branch of nonlinear solutions exists close to a bifurcation, strong non-normality can make the branch unreachable by time-stepping in practice, as 
mentioned 
in Ref.~\cite{gesla2024subcritical}.\\



\section{Nonlinear rolls dynamics}

\subsection{Pairing mechanism reinterpreted \label{sec:pairing}}

{The spatial localisation property of the eigenmodes, highlighted in fig.~\ref{fig:spectra}, has important consequences on the dynamics of the rolls, even in nonlinear unsteady regimes.}
Evidence for this is {analysed in detail} in the case $\delta=0.4$, a parameter for which the base flow first loses stability at $Re_c \approx 8,165$ in favour of the mode $B2$. The instability was found to be supercritical and the flow saturates nonlinearly, for $Re\gtrsim Re_c$,
 to a periodic flow.  

{As the primary eigenmode saturates in amplitude, the generation of harmonics of higher frequency leads to remarkable spatial dynamics.} \\

\begin{figure}
    \centering
    \includegraphics[width=0.49\linewidth]{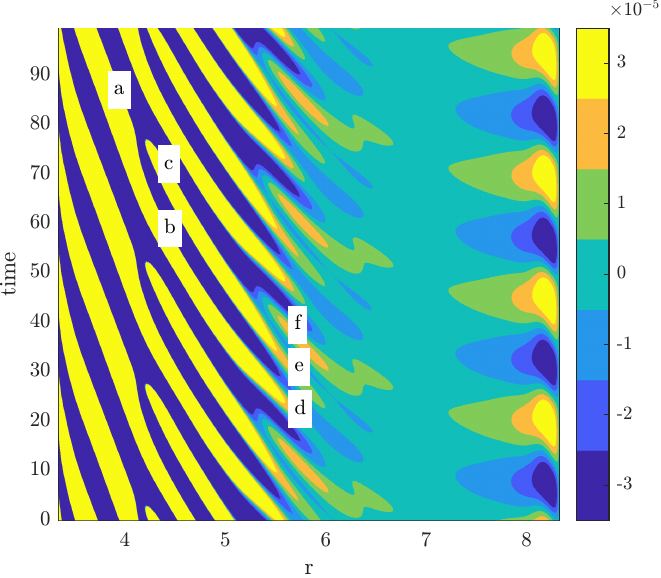}
    \includegraphics[width=0.49\linewidth]{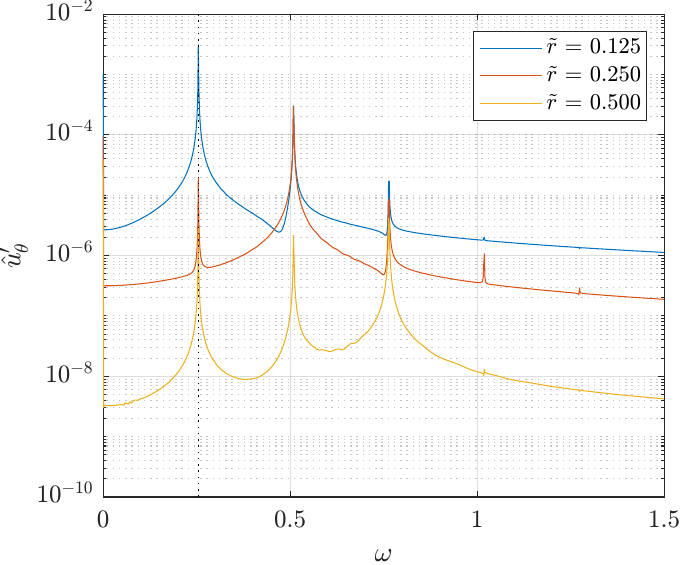}
    
    \caption{Space--time diagram of the fluctuation $(u_{\theta}-<u_{\theta}>)(r,z^*,t)$ at $z=z^*=0.94$, for $\delta=0.4$ and $Re=13,000$.  Left: snapshot of $u_{\theta}$ fluctuation velocity contour.
    \review{Going downstream the B\"odewadt layer, three  structures (d,e,f) merge into two (b,c) at \review{$r \approx 5.5$} while two structures (b,c) merge into one (a) at \review{$r \approx 4$}.}
    Right : {angular} frequency amplitude spectra from azimuthal velocity probes at three positions $r=3.96,4.58,5.83$, $\tilde r =(r-R_1)/(R_2-R_1)$.
    The dashed line marks the frequency of the primary mode $\omega \approx 0.2545$. 
    }
   \label{fig:sp_d04}
\end{figure}

\begin{figure}
    \centering
    \includegraphics[width=0.49\textwidth]{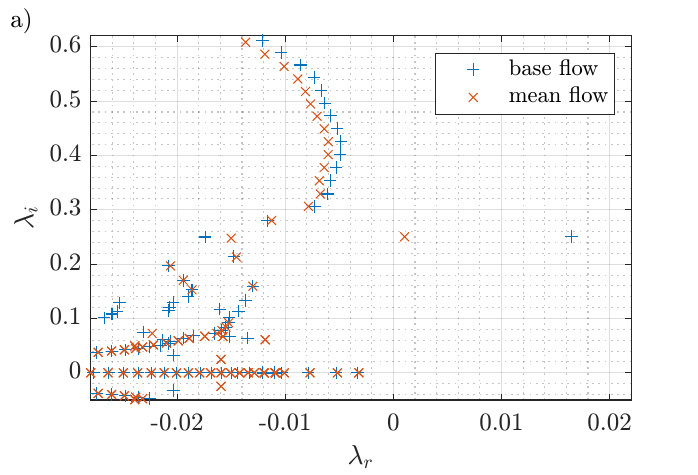}
    \includegraphics[width=0.49\textwidth]{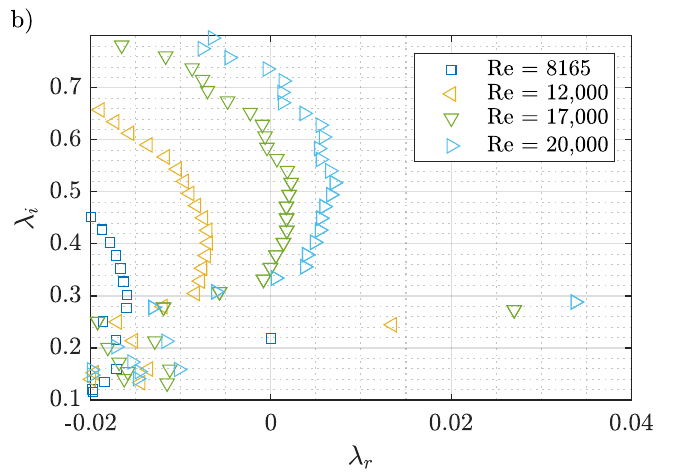}
    \caption{Spectrum of the Navier--Stokes operator linearised around the base flow and the 
    mean
    flow for Re=13,000 (a) and for different Re (b) at $\delta=0.4$. The real part of the most unstable eigenvalue is reduced where the linearisation point is switched from the base flow to the mean flow of nonlinearly saturate branch. This is consistent with the RZIF \review{(real-zero imaginary-frequency)} property. \review{Also note that the eigenvalues seem to lie on a continuous curve in some parts of the complex plane even though they are a discretised set due to the system being of finite length.}
    }
    \label{fig:rzif}
\end{figure}


Fig.~\ref{fig:sp_d04} shows a space-time diagram of the azimuthal velocity fluctuation along one given radial line inside the B\"odewadt layer (at $z=z^*$) for $Re=13,000$ where \emph{roll pairing} events can be observed.
At approximately mid-radius streaky structures are seen to propagate inwards, forming a front near $r \approx 6$. Between locations $r\approx6-8$ and $r\approx4.5-5.5$ the dominant angular frequency of the oscillations inside the front changes in a ratio 2/3 from $\omega \approx 0.76$ to $\omega= 0.51$. Moving further inwards down to $r \lesssim 4$, it decreases again to $\omega \approx 0.254$, one half of the angular preceding frequency, as found from  velocity spectra at various locations in~fig.\ref{fig:sp_d04}. We note that for this configuration the \emph{primary}, most unstable mode has an eigenvalue $0.0181 + 0.2545i$, which is exactly the angular frequency $\omega_1=0.2545$ identified closest to the axis in fig.~\ref{fig:sp_d04}. The dynamics of roll merging, as seen in fig.~\ref{fig:sp_d04}, can be explained as follows. The primary mode of angular frequency $\omega_1$  generates nonlinearly a second harmonic of angular frequency $2\, \omega_1$. {This oscillation excites, among the modes in the spectrum, one of the eigenmodes with angular frequency closest to $2\, \omega_1$, belonging to the continuous-like
branch visible \review{in fig.~\ref{fig:rzif}(a) for $\lambda_i \ge 0.3$.}. 
Because of its higher frequency, that eigenmode has its spatial support located upstream of the primary mode, as shown in fig.\ref{fig:spectra}. 
The spatial coexistence of two different frequencies is interpreted as roll pairing {and a pairing appears there where both modes are of comparable amplitude.} The story repeats itself through the nonlinear generation of a $3 \: \omega_1$ forcing which makes an eigenmode of frequency close to $3 \:\omega_1$ resonate.
Since the corresponding mode is located further upstream, a second pairing event appears.
At the location corresponding to (d,e,f) in fig.~\ref{fig:sp_d04}, one roll out of three disappears.
In general, for this phenomenological {start of a} cascade to become visible, the amplitude of the primary mode has to be large enough to induce second and third harmonics sufficiently large to make the corresponding modes visible. This explains why $Re$ has to be increased to a large supercritical value ($1.3\times 10^4$) well above $Re_c \approx 8,165$. We checked that, despite this large distance from criticality,
the corresponding base flow has still only one unstable mode (see fig.~\ref{fig:rzif}(a)).\\

The spectrum of the operator linearised around the \emph{mean} flow (computed by long time integration) was also computed, inspired by Ref.~\cite{barkley2006linear}. As seen in fig.~\ref{fig:rzif}(a), the mean flow is, in a first approximation, neutrally stable. It can thus be considered as belonging marginally to the class of flows sharing the RZIF \plq{\review{(real-zero imaginary-frequency)}} property~\cite{Turton2015prediction}, although the frequency of the mean flow is very little displaced from that of the base flow. Beyond that level of approximation, the non-strictly zero value of $\lambda_r$ is a signature of the non-negligible amplitudes associated with harmonics \cite{sipp2007global}.
Let us also mention that the 
visibility of this pairing mechanism is sensitive to the choice of boundary condition on the shroud. Pairing is more easily visible when the shroud
is attached to the stator as in the experiments by Schouveiler~\textit{et al.}~\cite{schouveiler1999spiral}, or with a linear azimuthal velocity profile as in the numerical simulations reported in~\cite{cousin1996thesis}.\\
We again emphasize  the crucial role of the radial dependence of the radial velocity in the  B\"odewadt layer in the reported phenomenology. Its key ingredients are the constant ratio between the phase velocity and the maximum velocity and the affine dependence of the velocity on radius. This is shown in fig.~\ref{fig:phase-speed} which results in the spatial localisation of the higher frequency eigenmodes illustrated in fig.~\ref{fig:spectra}. \\


\review{The phenomenology described above constitutes an alternative
\review{
} to the linear multi-harmonics forcing considered  in~\cite{gesla2024ontheorigin}. 
{Both mechanisms are an illustration of the pairing phenomenon~\cite{schouveiler2001instabilities,lopez2009crossflow}.}} They are however not in direct competition since one phenomenon occurs only at high $Re$ and for high amplitudes, whereas the other holds for all $Re$. The only difference between the two mechanisms concerns the way higher harmonics are generated~: in one scenario they are a consequence of the nonlinear term in the governing equations, whereas in the other the higher harmonics are explicitly introduced as part of a multi-harmonic forcing. 


\subsection{Coexistence of nonlinear regimes}

The time-periodic regime described above for $\delta=0.4$ was identified as the only attractor of the system for the parameters chosen. As $\delta$ is decreased, bistability is found, meaning that the attracting state depends on the initial condition. This is illustrated in fig.~\ref{fig:st_d012} 
for the case $\delta=0.12$, $Re=55,000$. For $\delta=0.12$, $Re_c=37,530$, {is} associated with a critical angular frequency of $0.208$; for $Re=55,000$ linearisation around the base flow yields only one unstable eigenvalue $0.013+0.232i$.
In the first row of fig.~\ref{fig:st_d012} a time-periodic regime, qualitatively similar to that for $\delta=0.4$, is uncovered. It also features a primary mode with angular frequency $\omega_1$ accompanied by two harmonics $2\,\omega_1$ and $3\,\omega_1$. However, pairing is not observed in the corresponding space-time diagram, which suggests that the amplitudes of these harmonics, although visible in the frequency spectrum, are not as strong as for the parameters in Subsection~\ref{sec:pairing}. \\

For strictly the same parameters, another attracting state is found by starting the simulation from a different initial condition \review{(resulting from perturbing the base flow with a random divergence free field of sufficient amplitude)}, shown in the lower row of fig.~\ref{fig:st_d012}. The corresponding frequency spectrum is broadband with no preferred frequency, the standard signature of deterministic chaos. Close to the hub, spectral amplitudes follow an exponential law.
For the most outwards radii, the frequency spectrum shows most of the energy concentrated near angular frequencies $\omega=0$ and $\omega \approx 1$, but unlike the supercritical case discussed above, no simple interpretation is terms of eigenmodes can be given. No dynamical origin for this chaotic state could be found. In analogy with the attractors found in linearly stable shear flows~\cite{eckhardt2018transition}, it is apparently disconnected from the base flow, meaning that it does not emerge from the amplitude saturation of an unstable eigenmode.\\ 

\begin{figure}
    \centering
    \includegraphics[width=0.49\linewidth]{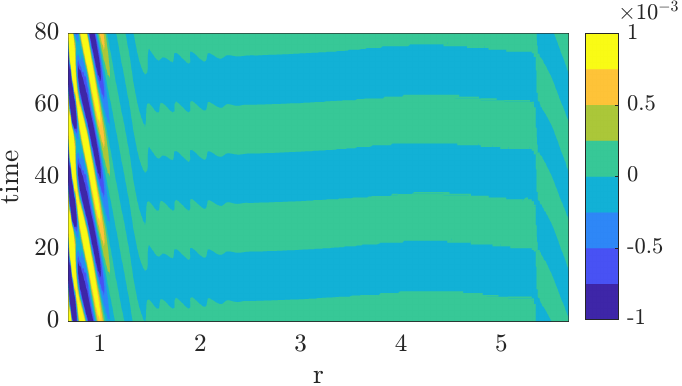}
    \includegraphics[width=0.45\linewidth]{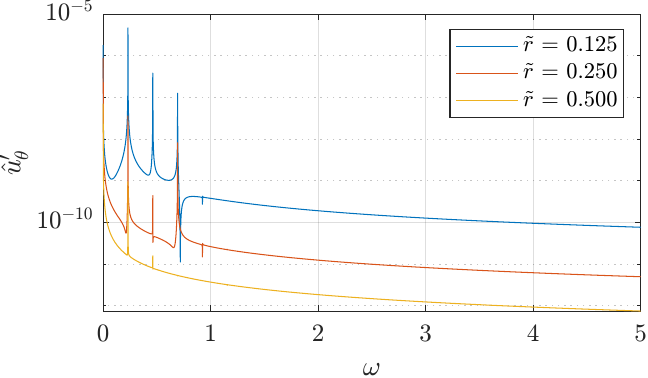}
    
    \includegraphics[width=0.49\linewidth]{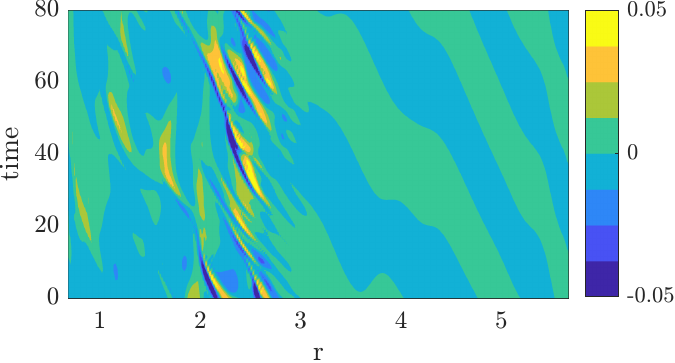}
    \includegraphics[width=0.45\linewidth]{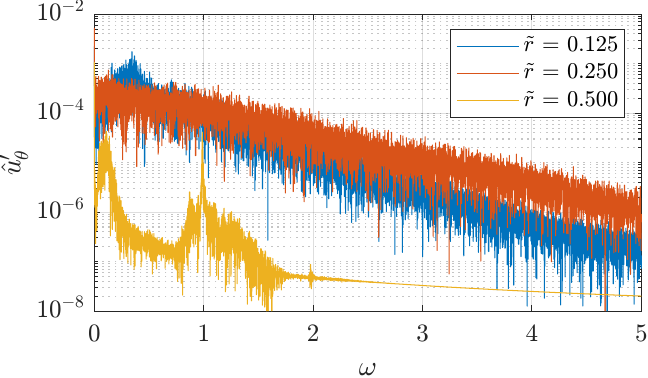}
    
    \caption{Space time diagrams (left) and corresponding spectra of probe signal (right) for $\delta= 0.12$ at $Re=55,000$ for the periodic solution (top) and the chaotic solution (bottom). It is noteworthy that two solutions exist at exactly the same value of the governing parameters $Re$ and $\delta$.
    Therefore, depending on the initial condition for the time integration, either of the two states can be reached.  
    For this configuration $Re_c=37530$ and the angular frequency of the neutral mode is $\lambda_i=0.208$. At $Re=55000$ this eigenvalue shifts to $\lambda=0.013\pm0.232i$. \review{Initial transients have been discarded.}
    } 
    \label{fig:st_d012}
\end{figure}

\subsection{Rotor stator case $\delta=0$.}

\subsubsection{Large-amplitude branch}


Our initial aim was to catch solutions for the case \review{$\delta  = 0$} belonging to the supercritical branch emanating from $Re_c$ by using the nonlinear timestepping code
\review{starting from the base flow perturbed with a random field of amplitude $10^{-10}$}.
By doing so we uncovered a disconnected branch of chaotic solutions that can be continued to $Re$ even below $Re_c$.\\

The evolution of the $L_2$ norm of the perturbation reported in fig.~\ref{fig:L2Vtheta} shows that, starting from an initial perturbation of amplitude $10^{-10}$ and after the initial non-normal growth, the norm of the perturbation decreases to $10^{-9}$ before it starts to increase exponentially with a growth rate that agrees with the linear stability analysis. After this phase of exponential growth the solution starts to grow faster than exponential and ends up eventually on a large amplitude branch characterised by unsteady chaotic fluctuations and a broadband frequency content (not shown). The fact that the solution ends up on a chaotic branch is not due to non-normal growth, rather to the fact that the supercritical branch emanating at $Re_c$, if any, folds back rapidly, and does not exist for  $Re= 6 \times 10^4$. This situation is exactly the same at that reported in~\cite{gesla2024subcritical} for $\Gamma =10$, where the supercritical branch was found to exist only for $ (Re - Re_c)  < 10^{-3}$ before folding back subcritically. 
Starting from this chaotic solution $Re$ was progressively decreased in order to locate the nose, in other words the tipping point of this large amplitude subcritical branch. It was found to lie in the interval $ 3.3 \times 10^4 < Re < 3.4 \times 10^4$. Surprisingly, just before this point, a periodic solution was identified for $Re= 3.4 \times 10^4$. 
Its period is \review{$T=18.6$}. The corresponding space-time diagram is shown in fig.\ref{fig:del0std}\review{(a)}. 
This periodic solution might be misinterpreted as belonging to a supercritical branch, which is clearly not the case since $Re$ is below $Re_c$ (see Table~\ref{tab:1}).\\


Comparing the time-periodic regime in fig. \ref{fig:del0std}\review{(a)} to the chaotic regime in 
fig. \ref{fig:del0std}\review{(b)} at slightly higher $Re$, we see that the rolls are closer to the axis  in the periodic regime. In the chaotic regime the propagation velocity fluctuates from roll to roll. 
Even closer to the axis for $1 \le r \le 2$ subharmonic frequency can be observed. This chaotic regime compares favourably to the one shown in fig. \ref{fig:st_d012} for non-zero $\delta$, yet higher $Re=55,000$, and without the subharmonic frequency reported at $\delta=0$ and smaller $Re$. As shown in Ref. \cite{gesla2024subcritical} for larger $\Gamma$, the maximum amplitude of the rolls corresponds to locations where fluid is ejected from the B\"odewadt layer towards the bulk. {Closer examination} of the velocity field suggests that these ejections contribute to the sustainment of the inertial oscillations in the bulk. }\\


\begin{figure}[ht]
    \centering
    \includegraphics[width=0.5\textwidth
    ]{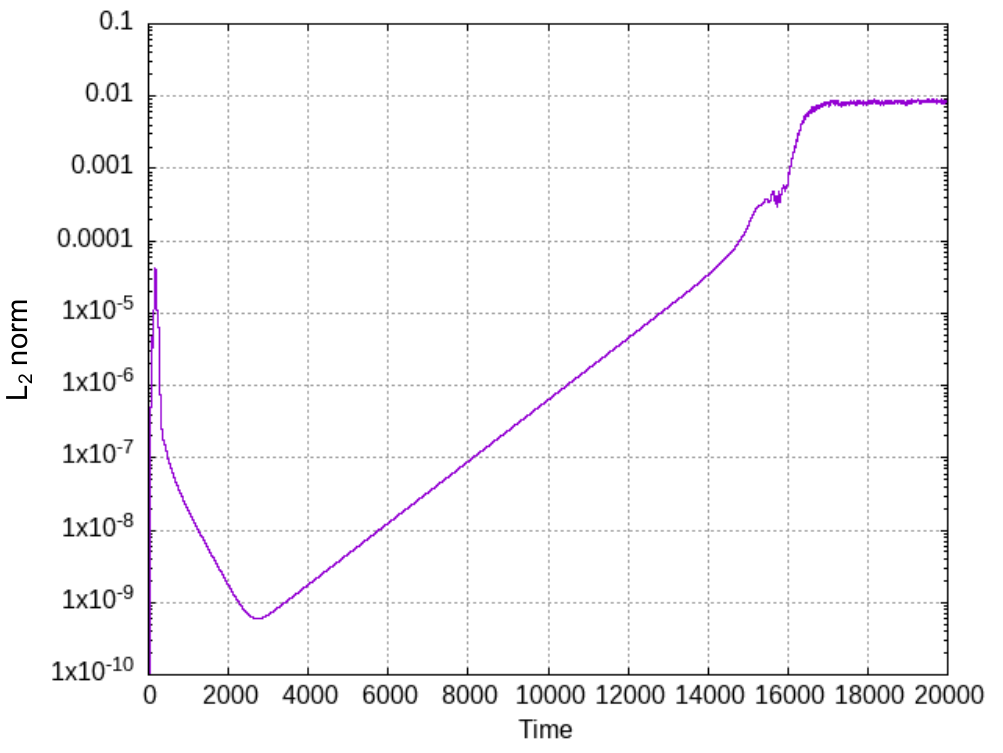}
    \caption{
    Time evolution of the $L_2$ norm of $v_\theta$; $ Re = 6 \times 10^4$, $\delta=0$.  
    }
    \label{fig:L2Vtheta}
\end{figure}

\begin{figure}
    \centering
    \includegraphics[width=0.49\textwidth]{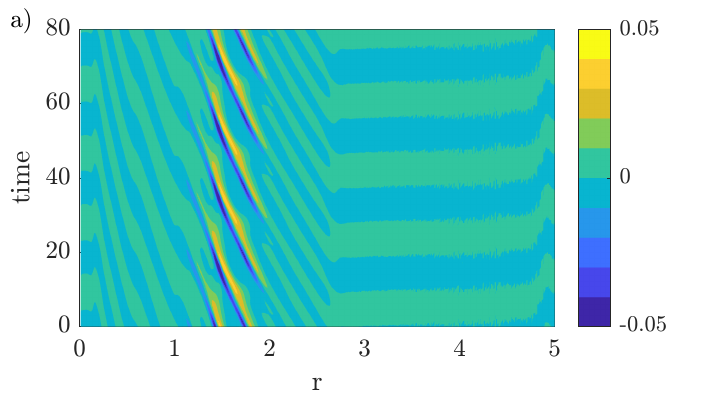}
    \includegraphics[width=0.49\textwidth]{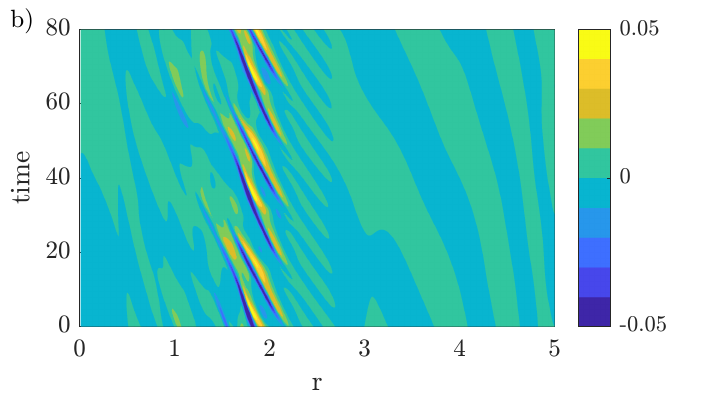}
    \caption{Space--time diagram of the fluctuation $(u_{\theta}-<u_{\theta}>)(r,z^*,t)$ for $\delta=0$ at $z=z^*=0.94$. {a) : periodic regime ($Re=34,000$), b) : chaotic regime ($Re=37,000$).}}
    \label{fig:del0std}
\end{figure}

\subsubsection{Edge state}

\begin{figure}[ht]
    \centering
    \includegraphics[width=0.49\textwidth, bb = 30 250 550 600]{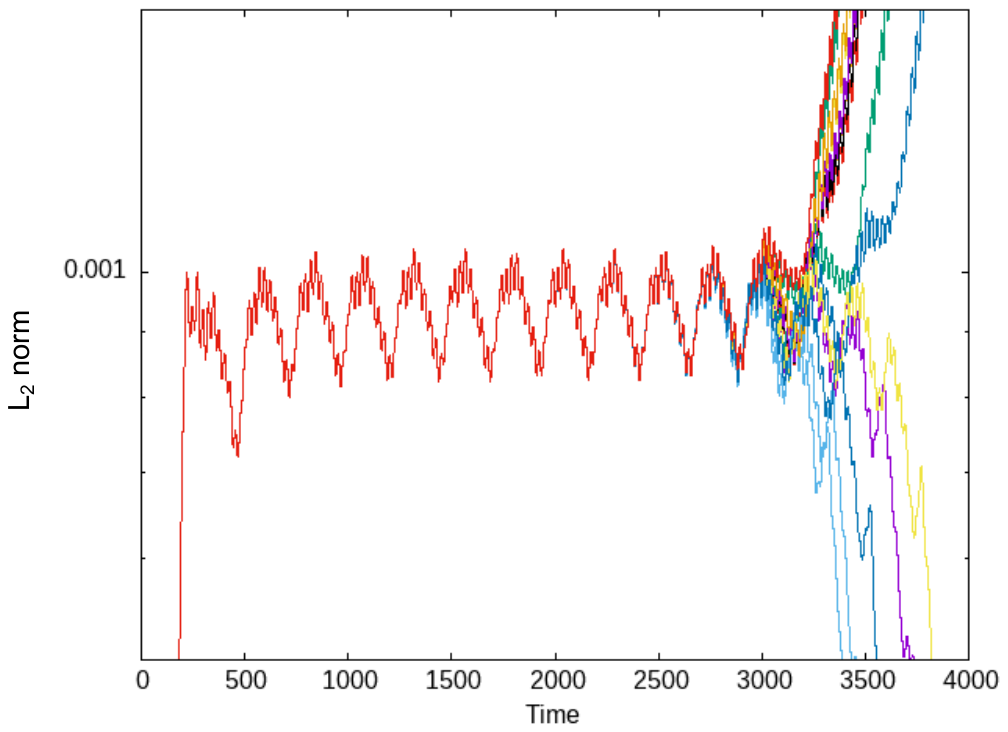}
    \includegraphics[width=0.49\textwidth, bb = 30 250 550 600]{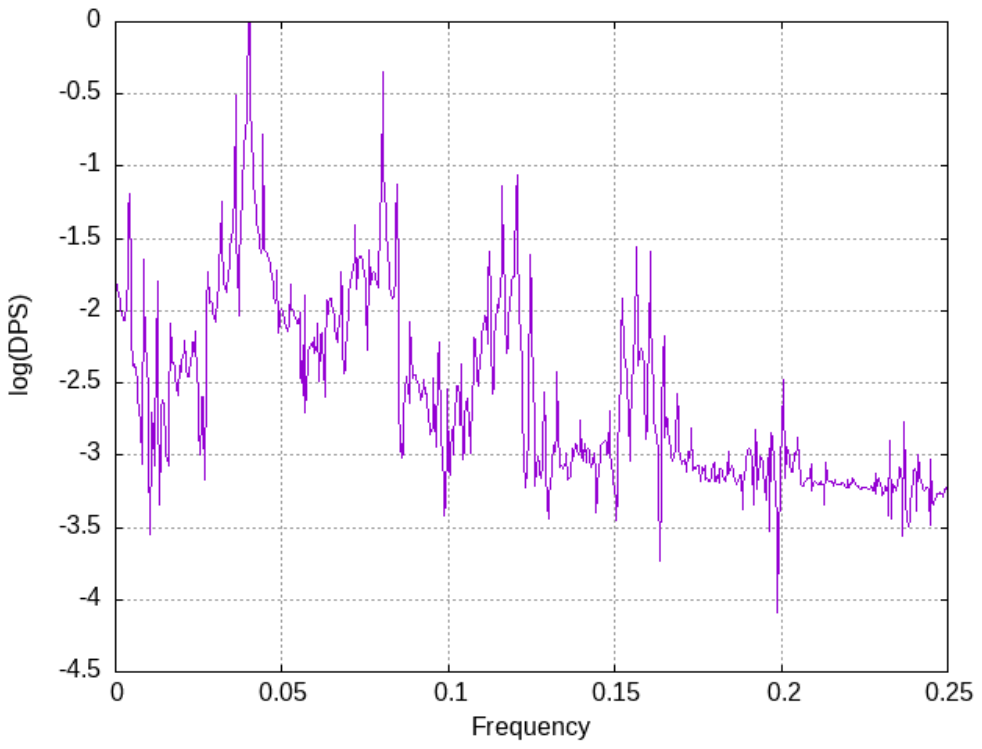}
    \caption{Trajectories bracketing an edge trajectory for $Re=5 \times 10^4 $, $\delta = 0$: (left) time evolution of the $L_2$ norm of fluctuating $v_\theta$ ; (right) power spectrum {of corresponding pointwise azimuthal velocity}, sample of 16384 points over time length of 1634.4 starting at t=1000. The trajectories shown approximate the edge state for $1000 \le t \le 2500$. $L_2$ norm of amplitude of initial noise is $7.6 \times 10^{-8} $.}
    \label{fig:edgere5e4}
\end{figure}

\begin{figure}[ht]
    \centering
    \includegraphics[width=0.49\textwidth, bb = 30 250 550 600]{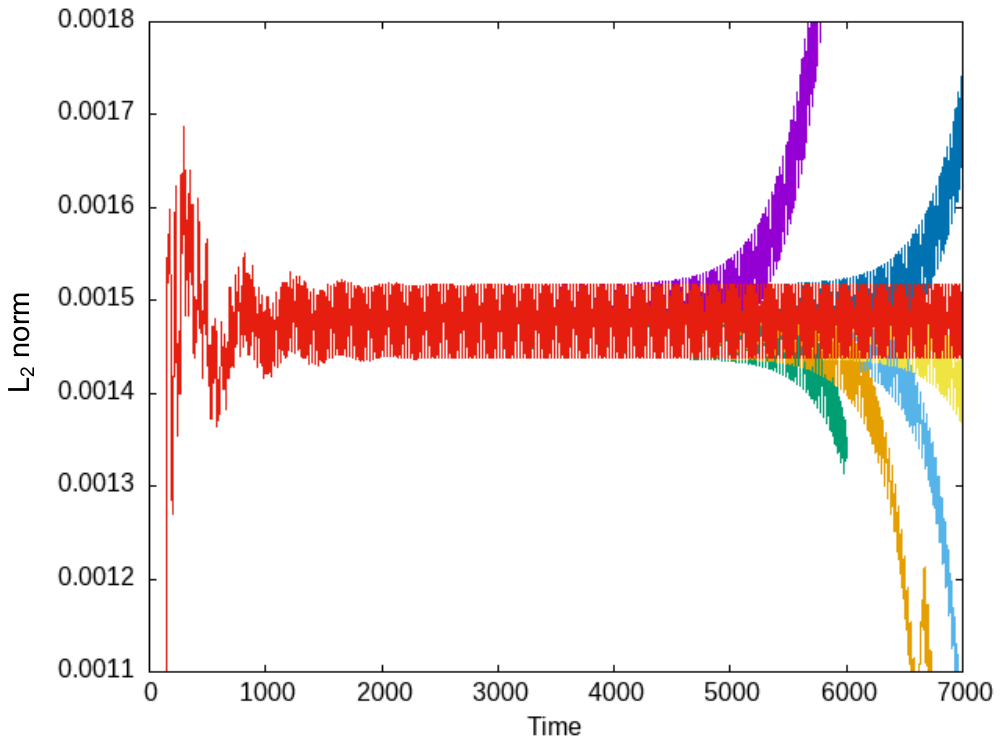}
    \includegraphics[width=0.49\textwidth, bb = 30 250 550 600]{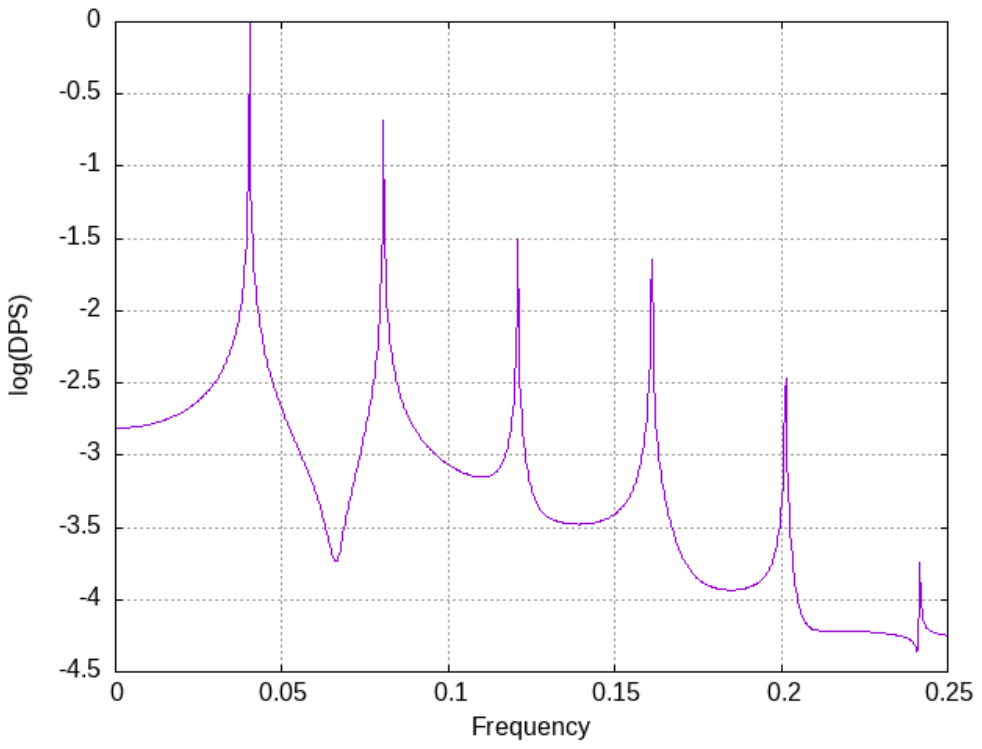}
    \caption{Trajectories bracketing an edge trajectory for $Re=4 \times 10^4$, $\delta = 0$~: (left) time evolution of the $L_2$ norm of fluctuating $v_\theta$ ; (right) power spectrum of corresponding pointwise azimuthal velocity, sample of 32768 points over time length of 3276.8.
    The trajectory in red approximates the edge state for $2000 \le t \le 7000$. $L_2$ norm of amplitude of initial noise is $1.9 \times 10^{-9} $.}
    \label{fig:edgere4e4}
\end{figure}

For the case $\delta=0$, additional nonlinear computations of \emph{edge states} have been performed, with the goal to list the most important finite-amplitude states for a bifurcation diagram to be drawn. 
Edge states~\cite{skufca2006edge} are unstable states attracting the trajectories constrained on the laminar basin boundary. They exist in typically subcritical conditions, including when the top branch is disconnected from the (linearly stable) base flow solution. 
These edge states are computed \review{here} starting from the base flow solution, to which a given and reproducible azimuthal velocity perturbation is added,
 whose amplitude is recursively tuned such that the simulation evolves neither towards the laminar base flow nor towards the chaotic state. For $Re=5 \times 10^4$ the edge state is quasi-periodic in time (fig.\ref{fig:edgere5e4}) with two incommensurate frequencies. This result is similar to the edge computation carried out for an aspect ratio of 10 \cite{gesla2024subcritical}. For $Re=4 \times 10^4$ the edge state, however, is strictly periodic in time (fig.~\ref{fig:edgere4e4}) with less intense fluctuations and a period $T \approx 24.8$.
These computations confirm the subcriticality of the state appearing at $Re_{SN} \approx 3.2 \times 10^4$. The bifurcation leading from the quasi-periodic to the periodic edge state might be of Neimark-Sacker type, or might correspond to synchronisation inside an Arnold's tongue, but this has not been investigated further.\\

\section{Bifurcation sequences : from supercritical to subcritical in $Re$}

We now compile here the lessons learned from the computation of $Re_c$ for all $\delta's$ and from a series of numerical explorations of nonlinear regimes at various parameter values. Our observations yield a qualitative description, as $\delta$ is reduced, of the sequences of bifurcations leading to unsteadiness as $Re$ increases.\\

\begin{figure}
    \centering
    \includegraphics[width=\linewidth]{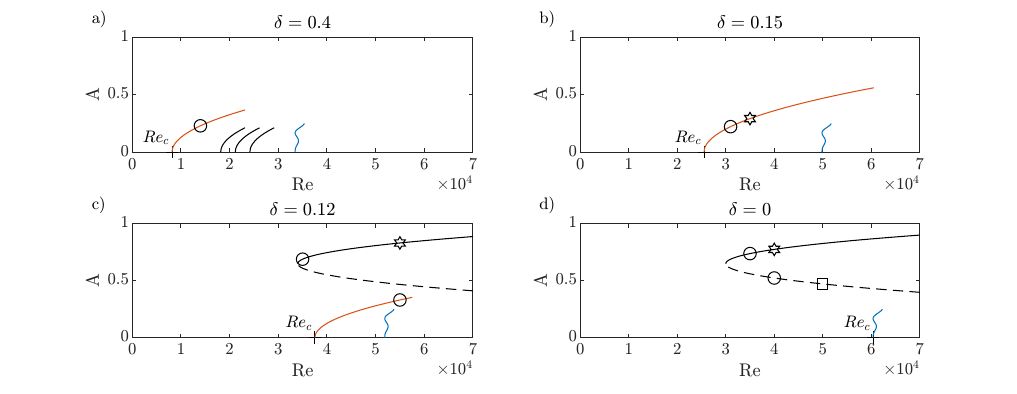}
    \caption{
     Sketch of the evolution of the bifurcation diagrams for selected values of $\delta$. Branches in red correspond to a B2 eigenmode while those in blue to a B1 eigenmode.  
     Additional symbols are confirmed solutions \review{($\circ$ : periodic, $\square$ : quasiperiodic ,
\fcStar{0.06}{black}{0.2}
     : chaotic), including} two edge states at $\delta=0$, \review{the} top branch state and the supercritical branch state at $\delta=0.12$, \review{and the} supercritical branch as in figure \ref{fig:sp_d04} at $\delta=0.4$).
    {The global transition picture switches from supercritical to subcritical as $\delta$ is decreased.}
    {The} amplitude $A$ is arbitrary and not to scale. \review{The blue line emerging from the B1 bifurcation has been intentionally shown as wavy in connection with the results published in Ref. \cite{gesla2024subcritical}}.
    }
    \label{fig:scenarios}
\end{figure}

\begin{figure}[H]
    \centering
\includegraphics[width=0.5\textwidth]{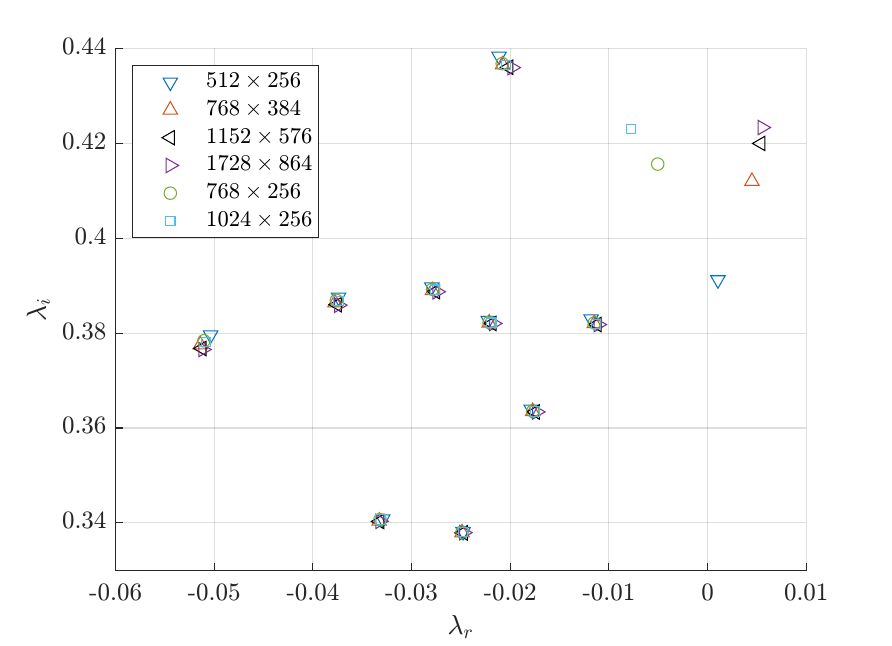}
    \caption{Sensitivity of the most unstable eigenvalues to the mesh resolution at $Re=60,000$, $\delta=0$. 
    }
    \label{fig:my_label}
\end{figure}

We chose to start the description of the various regimes at $\delta=0.4$ and to decrease $\delta$ from there towards the rotor-stator case at $\delta=0$. The different cases are summarized in fig.~\ref{fig:scenarios}.
\begin{itemize}
    \item At $\delta=0.4$, the transition to unsteadiness is supercritical : for $Re<Re_c$ the only stable state is the base flow while for $Re>Re_c$ the base flow is unstable and the system evolves towards the complex dynamics described in fig.~\ref{fig:sp_d04} (red branch in fig.~\ref{fig:scenarios}a) showing the ($\omega$,$2 \, \omega$,$3\,\omega$) pairing as described in  section~\ref{sec:pairing}. There are many secondary bifurcations points in between $Re_c(B2)$ and  $Re_c(B1)$.
    At $Re=1.5 \times 10^4$, the solution on the supercritical branch becomes quasiperiodic, and chaotic at larger $Re$.
\item For $\delta=0.15$, the most unstable branch emerging at $Re_c$ and based on $B2$ (red in fig.~\ref{fig:scenarios}b), still emerges supercritically, followed by a second destabilising branch (blue).  Note that the two corresponding values of $Re_c$ increase as $\delta$ decreases. 
Attempts to find a large amplitude branch for increasing values of $Re$ up to $7\times 10^4$ remained unsuccessful.
\item For $\delta=0.12$ the situation is relatively similar to $\delta=0.15$, except that the critical values of $Re$  have moved up in $Re$, and for the notable apparition of the disconnected branch at values of $Re$ comparable to $Re_c$. 
As a result, the complex dynamics on the upper branch is now in direct competition with the supercritical dynamics, both displayed in fig.\ref{fig:st_d012}. There is even a properly subcritical interval for $Re>35,000$ where the disconnected branch is in competition with the stable base flow below $Re_c=37784$. 

\item The case $\delta=0$ follows from the simple continuation to lower $\delta$ of the former cases. The critical value for the mode $B1$ has moved to a relatively high value  $Re_c \approx 6 \times 10^4$ while that for $B2$ has diverged.  
A direct search for supercritical solutions at slightly supercritical value of $Re$ lead instead to the large amplitude branch, confirming the result that this branch, if supercritical,
is virtually unreachable.
The large amplitude branch extends to values of $Re$ around $3.2 \times 10^4$ and ends in what appears to be a saddle node fold. Just before a time-periodic solution is found for $Re =3.5 \times 10^4$.  
Edge states are found on the border of the attraction basin between the base flow and the large amplitude branch. The edge state identified
($Re=4 \times 10^4$) close to the tipping is found to be periodic in time whereas that found for $Re=5 \times 10^4$ is biperiodic, just like for $\Gamma = 10 $~\cite{gesla2024subcritical}.
\end{itemize}

\section{Conclusions}

We have considered the axisymmetric flow inside an annular duct with a rotating disk, governed by a Reynolds number $Re$ and a geometric homotopy parameter $\delta$. In this system, the $\delta \rightarrow 0$ limit corresponds to axisymmetric {hub-free} rotor-stator flow while the $\delta \rightarrow 1$ limit corresponds to a flow analog to a two-dimensional differentially heated cavity.
A description of the transition scenario in $Re$ as $\delta$ decreases is given together with results from linear stability theory as well as with quantitative estimations of non-normal amplification. A plausible scenario {of minimal complexity} is suggested in Fig.~\ref{fig:scenarios}.
As $\delta$ decreases towards zero, the dynamics switches from supercritical to subcritical {in $Re$}, both because the threshold values of $Re_c$ for the destabilisation of global modes increase, and because the onset $Re_{SN}$ of the competing disconnected state occurs at lower and lower values of $\delta$. For low enough $\delta$, the amplification of disturbances due to non-normal effects is strong enough so that the supercritical branches are not reachable in practice in direct simulations.
The disconnected state becomes hence in practice the only attractor for $Re>Re_{SN}$.\\

 The observations compiled above lead to a more complete picture of the {axisymmetric} transition scenario for the rotor-stator case at $\delta=0$ with an aspect ratio of 5. The base flow loses its stability at a finite value of $Re_c \approx 6 \times 10^4$, 
 not identified
in previous numerical investigations of~\cite{lopez2009crossflow}, 
since it was 
restrained to lower values of $Re$. {Identifying this critical value would have been virtually impossible with the sole use of a nonlinear time-stepping code.} Nevertheless this linear instability and the steep supercritical branch emanating at $Re_c$ are relatively exotic~: non-normality will amplify almost any initial disturbance up to levels where they transition to the disconnected state, whether for $Re$ above or below $Re_c$, as soon as $Re \gtrsim Re_{SN}~( \approx  3.3 \times 10^4) $. This brings the phenomenology very close to that encountered in other shear flows such as pipe flow or plane Couette flow~\cite{eckhardt2018transition}. The disconnected state, together with the edge state as its unstable counterpart, was found to be either chaotic or time-periodic close to the tipping point, whereas the edge state was found either quasiperiodic or also time-periodic closer to the tipping point. 
If we compare now to the bifurcation diagram obtained recently in 
Ref.\cite{gesla2024subcritical} for a rotor-stator of aspect ratio 10, many similarities can be found, notably the subcriticality of the disconnected state, but also the large non-normal amplification observed at high $Re$ values comparable to $Re_c$ and the associated consequences. Several interesting differences are noted however, for instance finite lifetimes of the chaotic state were not identified in the current simulations.
 As another difference with the case $\Gamma=10$  we reported the occurrence of
coexisting time-periodic and chaotic solutions as well as of time-periodic states on both the upper stable and lower unstable disconnected branch. These periodic states should be amenable to numerical continuation using arclength techniques, {which is the subject of ongoing work}.
{It would be interesting to see which of the conclusions of the present study also apply to the description of the three-dimensional transition process.}

\section*{Acknowledgments}
\review{The authors would like to acknowledge the reviewers for careful reading of the manuscript and many useful suggestions.}


\section*{Appendix A}

\begin{table}[H]
  \centering
  \begin{tabular}{c|cc|cc}
   \multirow{2}{*} {$N_r \times N_z$ } &  \multicolumn{2}{c|}{$\delta = 0 $} &  \multicolumn{2}{c}{$\delta = 0.1 $}  \\
      & $Re_c$ & $\omega$  & $Re_c$ & $\omega$ \\ \hline
     512 $\times$ 256 & 59564 & 0.3924 &  47941 & 0.3942 \\
     768 $\times$ 384 & 58217.5 & 0.4052 & 46602 & 0.404 \\
      1152 $\times$ 576 & 57874.5 & 0.4116 & 46324.5 &  0.409 \\
    1728 $\times$ 864 & 57817  & 0.4146  &46260.5 & 0.4112 \\
    \end{tabular}

    \caption{Grid convergence $\delta=0$ B1 (left) and $\delta=0.1$ B1 (right). 
    }
    \label{tab:my_label}
\end{table}

\section*{Appendix B}  \label{appendixBC}

\begin{table}[H]
	\centering
	\begin{tabular}{|c|c|c|c|}
		\hline	
		$BC_{R1}$ & $BC_{R2}$ &  $Re_{c}$  &  $\lambda_i $ \\ \hline
		ROT & FIX & 4030 & 0.36 \\
		LIN & LIN & 5315 & 0.3565 \\
		STF & FIX & 5833 & 0.211\\
		STF & LIN & 6462 & 0.4229\\
		STF & STF & 7666 & 0.2255 \\
 		STF & ROT & 7572 & 0.2266\\
		FIX & ROT & 8020 & 0.2108 \\
		\hline
	\end{tabular}
	\caption{Critical Reynolds number $Re_c$ and angular frequency of most unstable eigenmode $\lambda_i$ for various velocity boundary conditions on the inner hub $BC_{R1}$ and outer shroud $BC_{R2}$~; $\delta = 0. 5$~; STF = stress-free~; ROT = rotating~; FIX = null velocity~; LIN = linear profile for the azimuthal velocity as a function of $z$.  The reported values of $Re_c$ show that the most stable configuration corresponds to a steady hub and rotating shroud whereas the least stable is the rotating hub and steady shroud, with a factor two in $Re$ between these two extremes. } 
	\label{table:bcs}
	
\end{table}

\bibliography{biblio}



\end{document}